\documentclass[11pt,twoside]{book}
\usepackage{konkolyproc2}
\usepackage{longtable}
\usepackage{amsmath,amssymb}
\usepackage{graphicx}
\usepackage{lscape}
\usepackage{index}
\usepackage{natbib}
\makeindex

\begin{document}

\pagestyle{myheadings}
\setcounter{equation}{0}\setcounter{figure}{0}\setcounter{footnote}{0}
\setcounter{section}{0}\setcounter{table}{0}\setcounter{page}{1}
\markboth{Kov\'acs}{The Blazhko phenomenon}
\title{The Blazhko phenomenon}
\author{G\'eza Kov\'acs}
\affil{Konkoly Observatory, Budapest, Hungary}
%
\begin{abstract}
In spite of the accumulating high quality data on RR~Lyrae stars, the 
underlying cause of the (quasi)periodic light curve modulation (the 
so-called Blazhko effect) of these objects remains as mysterious as it 
was more than hundred years ago when the first RR~Lyrae observations 
were made. In this review we briefly summarize the current observational 
status of the Blazhko stars, discuss the failure of all currently 
available ideas attempting to explain the Blazhko effect and finally, 
we contemplate on various avenues, including massive 2-3D modeling to 
make progress. Somewhat unconventionally to a review, we present some 
new results, including the estimate of the true incidence rate of the 
fundamental mode Blazhko stars in the Large Magellanic Cloud, and tests 
concerning the effect of the aspect angle on the observed distribution 
of the modulation amplitudes for Blazhko models involving nonradial modes. 
\end{abstract}

\vspace*{-6mm}
%
%
\section{Introduction}
Cepheids and RR~Lyrae stars have always played an invaluable role in 
mapping the Universe and tracing the kinematical and physical properties 
of various stellar populations. With the advance of the Gaia mission, 
in a few years' time we will be able to fix the zero points of the 
period-luminosity-color or period - near infrared magnitude relations with 
an error lower than $0.01$~mag \citep{windmark2011} and thereby measure 
distances in the nearby Universe with an unprecedented accuracy. Although 
the physics behind the basic properties (such as their self-excitation 
mechanism) of these vital objects are well known for more than half of 
a century, there is a nearly complete lack of understanding two, 
well-populated subgroups of these variables. These subgroups are those of 
the double-mode pulsators (both among Cepheids and RR~Lyrae stars) and 
the Blazhko variables (characteristically among RR~Lyrae stars). Although 
the modes occurring in the `classical' double-mode variables are 
well-identified with the low-order modes from the linear 
theory\footnote{Although this statement is true for the fundamental/first 
overtone pulsators, the nature of the newly discovered class with period 
ratios of $\sim0.60$--$0.64$ is a mystery \citep[see][and 
references therein]{netzel2015}.}, the clear cause of the sustained double-mode 
state is still unknown and nonlinear modeling is controversial 
(see \citealt{kollath2002} vs. \citealt{smolec2010}). The situation with the 
Blazhko stars (RR~Lyrae variables showing periodic amplitude and phase 
modulations) is even worse. As of this writing, {\em we do not have a clue} 
why many RR~Lyrae stars vary their amplitudes that leads in some cases nearly 
ceasing pulsation in the low-amplitude states. This is not just a minute 
wrinkle spoiling the classical and simple picture on these old stars. On the 
opposite, with no physically justified and testable idea/model we miss some 
basic ingredient not just in the pulsation models but most likely in the 
evolutionary models, too. 

No detailed accounts are to be given in this summary on the observational 
properties and failed modeling of these objects. On these we refer to the 
review by \cite{kovacs2009} from the pre-Kepler era and those by 
\cite{kolenberg2011} and \cite{szabo2014} more recently, already incorporating 
the results of the Kepler mission. We list some obvious (but strenuous) 
avenues for future works, including detailed spectroscopic studies and 
higher dimension hydrodynamical surveys of RR~Lyrae models. Unconventionally, 
we present some new results concerning the simple consequences of the 
observed distributions of modulation amplitudes. 

\vspace*{-2mm}
%
%
\section{How many are they?}
Here we focus on the fundamental mode (RRab) variables, since their 
considerably larger number and higher amplitudes yield statistically 
more reliable samples than those available for the first overtone 
stars. The basic rates and related data are summarized in Table~\ref{bl_stat}.   

%
%
\vspace*{-3mm}
\begin{table}[!hb] 
\caption{Current incidence rates for RRab BL stars}
\smallskip
\begin{center}
\begin{tabular}{lrrl}
\tableline
\noalign {\smallskip} 
System & Rate & \# stars & Note\\ 
\noalign{\smallskip}
\tableline
\noalign{\smallskip}
Galactic field &  5\% & 1435  & ASAS; \cite{szczygiel2007}$^1$ \\
               & 47\% & 30    & ground-based; \cite{jurcsik2009} \\ 
               & 39\% & 44    & {\it Kepler}; \cite{szabo2014}$^2$ \\ 
               & 60\% & 13    & {\it CoRoT}; \cite{szaboetal2014} \\ 
               & 34\% & 268   & ASAS \& WASP; \cite{skarka2014} \\
Galactic bulge & 23\% & 215   & OGLE-I; \cite{moskalik2003} \\
               & 25\% & 1942  & OGLE-II; \cite{mizerski2003} \\
               & 30\% & 11756 & OGLE-III; \cite{soszynski2011}$^3$ \\
M3             & 50\% & 200   & ground-based; \cite{jurcsik2014} \\
LMC            & 12\% & 6135  & MACHO; \cite{alcock2003} \\
               &  8\% & 478   & OGLE-III; \cite{chen2013}$^4$ \\
               & 20\% & 17693 & OGLE-III; \cite{soszynski2009}$^3$ \\
SMC            & 22\% & 1933  & OGLE-III; \cite{soszynski2010}$^3$ \\
\noalign{\smallskip}
\tableline
\noalign{\smallskip}
\end{tabular}  
\end{center}
\scriptsize
\vspace*{-3mm}
 Notes:\\
 $^1$ The low incidence rate is most probably accounted for by the lower 
 sampling rate and the higher noise in the then available release of the 
 ASAS database. \\ 
 $^2$ See \cite{benko2014} for a slightly lower incidence rate. \\
 $^3$ No details are given. \\
 $^4$ Strange low incidence rate in spite of the high-quality data selection.  
\label{bl_stat}
\end{table}
\vspace*{-2mm}

What is clear from this table is that Blazhko (BL) stars are numerous in 
various stellar populations and the actual figures suggest internal 
differences among them (e.g., LMC vs. Galactic field). It is interesting 
that traditional ground-based surveys yield rates close to the ones derived 
by the orders of magnitude more accurate space 
missions.\footnote{Note that the Konkoly BL survey by Jurcsik and co-workers 
was based on the observations made by a $60$~cm telescope, whereas wide-field 
transit/variability surveys operate $\sim10$--$20$~cm-class telescopes.} This 
suggests that {\em the relative number of BL stars with low modulation 
amplitudes is small}. Indeed, from the $15$ BL stars of the {\it Kepler} sample 
discussed by \cite{benko2014} only two show modulation amplitudes lower than 
$0.01$~mag. This is a remarkable property, indicating the existence of some 
lower floor of the amplitude modulation and suggesting that reliable 
statistics can be derived also from the today's ground-based surveys capable 
of hitting the detection limit for sinusoidal signals with amplitudes of 
$\sim 0.005$--$0.01$~mag.

%
\subsection{The debiased incidence rate in the LMC}
We see from Table~\ref{bl_stat} that the incidence rates in both Magellanic 
Clouds are systematically lower that those derived in the Galactic bulge and, 
especially, in the Galactic field (GF). There is also a significant difference 
between the MACHO and OGLE rates. Since both rates are based on fairly large 
samples, we suspect that the difference is due to the higher accuracy of 
the OGLE data as they evolved over the years after the MACHO project was 
abandoned. In the following we correct the MACHO statistics for the lower 
detection rates and examine if it leads to an agreement with the OGLE 
rates. 

First we compute the Cumulative Distribution Function (CDF) of the maximum 
Fourier side lobes available from the published data (\citealt{alcock2003} [LMC];  
\citealt{jurcsik2014} [M3] and \citealt{skarka2014} [GF]). Recall that CDF 
denotes the probability that the modulation amplitude is smaller than a given 
value, i.e., $\rm CDF(MAX_{sidelobe})=P(A<MAX_{sidelobe})$. Figure~1 shows the 
resulting CDFs, clearly indicating a want of the low-amplitude BL stars in 
the LMC sample. With similar observational noise, the BL stars in M3 and in 
the GF seem to follow the same distribution (we do not deal with the small 
differences at this stage). 

%
\vspace*{-4mm}
\begin{figure}[h]
\begin{minipage}[t]{0.40\textwidth}
\includegraphics[angle=-90,width=1.30\textwidth]{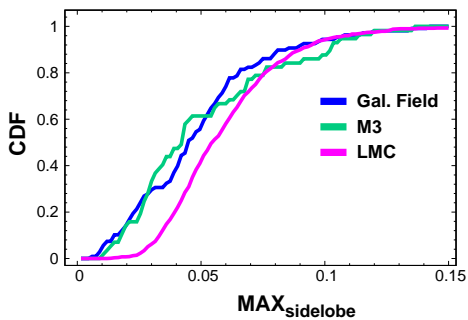}
\end{minipage}\hfill
\begin{minipage}[t]{0.60\textwidth}
\caption{\small Cumulative distribution functions of the largest sidelobes 
of the Fourier spectra for fundamental mode Blazhko stars in three different 
stellar populations.}
\end{minipage}
\label{cdf_obs}
\end{figure}
\vspace*{-4mm}

We test the hypothesis that BL stars both in the LMC and in the GF follow 
the same intrinsic CDF and the difference is entirely attributed to the 
lower detection efficiency on the noisier MACHO data. In a simple approach 
the test can be performed by injecting sinusoidal signals in the time series 
consisting the Gaussian noise, generated according to the standard deviations 
of the residuals (particular to each star) in the single mode RR~Lyrae sample. 
Instead of generating and then analyzing these time series, we estimate the 
signal-to-noise ratio (SNR) of a sinusoidal component with amplitude $A$, data 
point number $N$ and noise standard deviation $\sigma$   
%
%
\begin{equation}
{\rm SNR} = {A\sqrt{N/2} \over \sigma} \hskip 2mm . 
\end{equation}
We tested the applicability of this formula by comparing the above parameter 
with the S/N of the frequency spectra. We found good correlation, and set 
the lower limit of SNR to $5.5$ to fix the detection rate equal to the one 
obtained directly from the frequency spectra.   

%
\vspace*{-4mm}
\begin{figure}[h]
\begin{minipage}[t]{0.40\textwidth}
\includegraphics[angle=-90,width=1.30\textwidth]{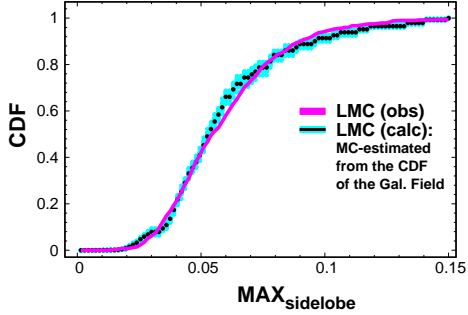}
\end{minipage}\hfill
\begin{minipage}[t]{0.60\textwidth}
\caption{\small {\em Light dots:} Monte Carlo-based cumulative distribution 
function of the modulation amplitudes for the LMC, following the noise 
properties of the MACHO~data base with the priors given by the 
observed distribution of the Galactic field Blazhko stars; 
{\em Black dots:} Averages of these realizations; {\em Continuous line:} 
Observed CDF of the RRab BL stars in the LMC.}
\end{minipage}
\label{field2lmc}
\end{figure}
\vspace*{-5mm}

After generating amplitudes following the CDF of the GF BL stars, many 
realizations are tested by using the standard deviations of the residuals 
in some $2000$ monoperiodic stars from the MACHO LMC sample. Each star 
is checked for the detection criterion ${\rm SNR}>5.5$ and flagged as 
`detected' or `not detected' accordingly. Since the amplitudes are known, 
we are able to construct the CDF for the detected cases. The result is 
shown in Fig.~2. We see that using the noise properties of the MACHO data 
on the CDF of the GF BL stars, the resulting CDF becomes very similar to 
the observed CDF of the LMC. This suggests that {\em the distribution of 
the modulation amplitudes is likely the same in the two populations}.    

%
\vspace*{-3mm}
\begin{figure}[h]
\begin{minipage}[t]{0.40\textwidth}
\includegraphics[angle=-90,width=1.30\textwidth]{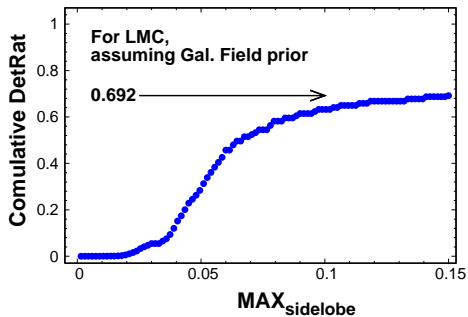}
\end{minipage}\hfill
\begin{minipage}[t]{0.60\textwidth}
\caption{\small Cumulative detection rate of sinusoidal signals on the 
MACHO/LMC data with the prior amplitude distribution as given by the 
Galactic field stars.}
\end{minipage}
\label{lmc_rat}
\end{figure}
\vspace*{-5mm}

The true incidence rate for the LMC can be calculated from the unnormalized 
version of the CDF based on the amplitude distribution of the GF (black 
dots in Fig.~2). This cumulative detection rate (see Fig.~3) yields a 
factor of $0.69$ for the total detection rate. This, with the observed 
rate of $12$\% of \cite{alcock2003} results in an unbiased incidence rate 
of $17$\%. Considering the values listed in Table~\ref{bl_stat}, we see a 
better (though not perfect) agreement with the values obtained from the more accurate OGLE 
data for both Clouds. This test supports the idea that {\em the lower 
incidence rate in the Magellanic Clouds is real and that there is a likely 
population dependence of the occurrence of BL stars}.

%
\subsection{Modeling the distribution of the modulation amplitudes}
With the recent sample of the GF BL stars by \cite{skarka2014} and the 
already existing large amount of data on the Magellanic Clouds, we are 
in a position to test a simple BL model involving the $(l,m)=(1,\pm1)$ 
nonradial components. We recall that in the basic model of the BL effect 
this nonradial mode is the one that is most viable for a $1:1$ resonant 
interaction with the fundamental radial mode \citep[see][]{vanhoolst1998}. 
The coupling gives rise to a steady (constant amplitude) triple-mode 
pulsation with the $l=0$ radial fundamental, and the  $l=1$, $m=-1$ and 
$m=+1$ nonradial modes \citep[see][]{nowakowski2001}. The interaction leads 
to a phase-locked, that is single-period pulsation. The amplitude modulation 
is due to the changing sky-projected area with the nonradial component of 
the rotating star. Although in this simple form this model is unable 
to explain asymmetric modulation side lobes, it is still interesting how 
the observed distribution of the modulation amplitudes look like if the 
above mode pattern played some role in the BL phenomenon.   

Following \cite{dziembowski1977}, the observed luminosity variation of 
a nonradially pulsating star can be written in the following form 
(the abbreviated form of his Eq.~6) 
%
%
\begin{equation}
\Delta M_{\rm bol} = A_l^mP_l^m(\cos \theta_0)\sin(\omega_l^m t + \phi_l^m) \hskip 2mm . 
\end{equation}
Here $A_l^m$, $\omega_l^m$ and $\phi_l^m$ are, respectively, the surface amplitude, 
frequency and phase of the nonradial mode. The factor $P_l^m(\cos \theta_0)$ is 
the associated Legendre polynomial, depending on the mode order, degree and 
the inclination angle $i=\theta_0+\pi/2$ ($\theta_0=0$ if the rotational axis 
is perpendicular to the line of sight). In the simple case of the $(l,m)=(1,\pm1)$ 
mode, this factor is just a sine function. Below we examine the effect of this 
factor on the observed distribution of the amplitude modulation, assuming 
different priors for the distributions of the underlying nonradial mode 
amplitudes and a uniform prior for inclination angle.  

%
\vspace*{-4mm}
\begin{figure}[h]
\begin{minipage}[t]{0.40\textwidth}
\includegraphics[angle=-90,width=1.30\textwidth]{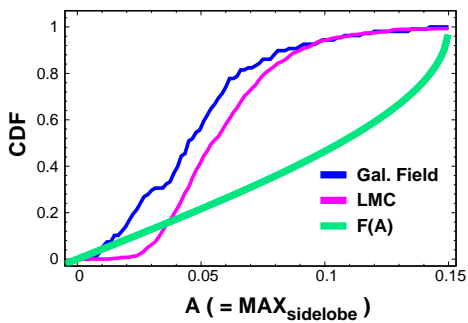}
\end{minipage}\hfill
\begin{minipage}[t]{0.60\textwidth}
\caption{\small Predicted CDF of the observed modulation amplitudes assuming 
Dirac delta prior (i.e., the same value) for the ($l=1$, $|m|=1$) nonradial 
component in all BL stars. The theoretical CDF can be computed analytically 
in this case, yielding $F(A)={2 \over \pi} \arcsin({A \over A_{\rm max}})$, 
where $A_{\rm max}=0.15$.} 
\end{minipage}
\label{orp_test1}
\end{figure}
\vspace*{-4mm}

First we assume that all BL stars have the same modulation amplitude and 
the observed amplitude is merely affected by the aspect angle. This, 
admittedly not too likely scenario leads to the theoretical CDF shown in 
Fig.~4. It is comforting that we can clearly exclude the  
extreme mechanism that might lead to such a particular amplitude distribution. 
Next, we consider the physically more likely setting, when all amplitudes 
in $[0, A_{\rm max}]$ occur with the same probability (here $A_{\rm max}$ 
is the maximum possible modulation amplitude -- which, as noted earlier, 
we take as the maximum sidelobe in the Fourier spectra). As shown in 
Fig.~5, this is a better-fitting model but still shows 
characteristic deviations by under/over-estimating the observed distribution 
at high/low modulation amplitudes. 

Finally, we assume that low/high modulation amplitudes are intrinsically 
of low probability. This situation is modeled by a Gaussian distribution 
peaked at $A=0.07$. We see on the right panel of Fig.~5 that this assumption 
yields a considerably better fit, although the high-amplitude part of the 
distribution is now overestimated. From these results we conclude that 
even if the BL phenomenon is affected by an aspect-dependence, there 
should exist some mechanism that prefers mid-size modulations and makes 
low/high modulations much less likely.    

\vspace*{-3mm}
\begin{figure}[h]
\begin{minipage}[t]{0.40\textwidth}
\includegraphics[angle=-90,width=1.40\textwidth]{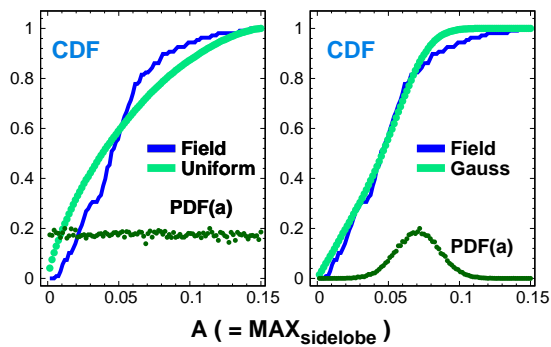}
\end{minipage}\hfill
\begin{minipage}[t]{0.53\textwidth}
\caption{\small {\em Left:} Predicted CDF (thick line) of the modulation 
amplitude, assuming uniform priors for the ($l=1$, $|m|=1$) nonradial 
component. For reference, the probability density (arbitrarily shifted) of 
these amplitudes generated in the Monte Carlo simulations are shown by the 
points labelled as PDF(a). {\em Right:} As on the left but for a Gaussian 
prior, chosen to fit better the observed distribution.}
\end{minipage}
\label{orp_test2}
\end{figure}

\vspace*{-8mm}
%
%
\section{Properties to be explained and the failure to do so}
There are many intriguing properties of the BL stars and it is obviously  
impossible to list them all here. What might perhaps still be useful is 
to focus on the more general, robust properties, that are inescapable to 
deal with in any future modeling. We list these properties in 
Table~\ref{bl_prop} with some sort of ranking based on a mixture of the 
size and/or the commonality of the given property. We do not know which 
feature will finally unveil the secret of the BL phenomenon but at this 
stage our preference goes to those ideas/models that offer some solution 
to reproduce the first three/four properties.    

%
%
\begin{table}[h] 
\caption{Basic properties of the RRab BL stars}
\smallskip
\begin{center}
\begin{tabular}{rlr}
\tableline
\noalign {\smallskip} 
\# & Property & Reference$^a$ \\ 
\noalign{\smallskip}
\tableline
\noalign{\smallskip}
1 & High ($20$--$50$\%) incidence rate & these proceedings \\
2 & Cases of high modulations ($A_{\rm mod}/A_{\rm puls}>0.5$)  & \cite{sodor2012} \\ 
3 & High rates of strongly asymmetric side lobes$^b$ & \cite{alcock2003} \\
4 & Range of modulation time scales ($\sim 5$--$1000$~d) & \cite{benko2014} \\
5 & Intermittent amplitude alternation$^c$ & \cite{szabo2010} \\
6 & Role of the 1st overtone and multimodality & \cite{smolec2015} \\
7 & Occurrence of multiperiodic modulations & \cite{skarka2013} \\
8 & Chaotic/stochastic effects? & \cite{plachy2014} \\
\noalign{\smallskip}
\tableline
\noalign{\smallskip}
\end{tabular}  
\end{center}
\scriptsize
\vspace*{-3mm}
 Notes:\\
 $^a$ The references are not complete. They are merely shown for guidance. \\ 
 $^b$ By the loose term `strongly asymmetric' we mean those cases when 
 one of the side lobes is close to the noise level. The number of these 
 stars may also depend on the stellar population (i.e., for the LMC this 
 ratio is $50$\%, whereas for the GF it is $40$\% -- see \citealt{skarka2014}). 
 At lower noise level they may exhibit both side lobes but they remain 
 quite different.\\ 
 $^c$ The phenomenon is commonly called in the RR~Lyrae community as 
 `period doubling'. We think that unless it can be clearly related to the 
 first step on the route of bifurcation to chaos, it is more appropriate 
 to call `amplitude alternation', since the expression `period doubling' is 
 specifically attached to the process mentioned \citep{feigenbaum1983}. 
 We also note the systematic frequency displacements at the positions of 
 the expected half integer resonances in some well-studied cases 
 \citep{bryant2015} are also atypical of period doubling.   
\label{bl_prop}
\end{table}

For the currently available models/ideas, here is a brief summary of the 
main reasons of their failure in their present forms.\footnote{We list  
only the most striking deficiencies of the models/ideas and refer to 
earlier reviews for a more extended discussion.} 

$\blacksquare$
{\em Magnetic oblique rotator/pulsator:} \cite{shibahashi2000}: 
symmetric sidelobes only, lack of strong magnetic field.  

$\blacksquare$
{\em Nonresonant radial/nonradial double-mode pulsator:} \cite{cox2013}: 
questions about the strength of the excitation of the nonradial mode 
and about the reasons why the close radial and nonradial modes are 
not phase-locked, leading to monoperiodic pulsation.  

$\blacksquare$
{\em Resonant nonradial rotator/pulsator:} \cite{nowakowski2001}: 
symmetric sidelobes only, questions concerning the current forms 
of amplitude equations in the case of nonradial modes in RR~Lyrae 
models \citep{nowakowski2003}. 

$\blacksquare$
{\em Radial mode resonance of 9:2 :} \cite{buchler2011}: lack of 
modulation in the RR~Lyrae nonlinear hydrodynamical models. 

$\blacksquare$
{\em Magnetic dinamo-driven convection:} \cite{stothers2011}: lack of 
mathematical/physical rigor and any modeling\footnote{See \cite{smolec2011} 
and \cite{molnar2012} for some negative hydrodynamical tests of this idea.}. 

$\blacksquare$
{\em Periodic energy dissipation driven by shock wave dynamics:} 
\cite{gillet2013}: the idea is based on the analysis of standard 
1D hydrodynamical models showing no amplitude modulation, therefore 
it is unclear what kind of mechanism could be deciphered from these 
models that might be relevant for future modeling.  

\vspace*{-3mm}
%
%
\section{Future progress: Can further observations or 3D modeling help?} 
Because of the lack of any, physically sound idea, it is hard to point 
toward any direction in which progress can be made. Nevertheless, 
we can mention two areas that are still not investigated with the depth 
required for getting useful information for model building. From the side 
of the observations, it would be vital to find out if there is any nonradial 
component. If yes, then does its size correlate with the amplitude of the 
Blazhko effect? It is a hard observational project (faint objects, long-term 
data acquisition, need for high-dispersion spectroscopy, difficulties in 
disentangling the nonradial component in the presence of the large radial, 
time-dependent component, etc.). From the side of modeling -- again, due to 
the lack of any better guidance -- one may try to conduct some expensive 
survey of the already existing 2D/3D hydrodynamical models (e.g., 
\citealt{mundprecht2015} and \citealt{geroux2015}), and investigate the 
effect of better modeled convection or that of the rotation. Of course, 
due to the well-known attribute of serendipity of scientific discoveries, 
it might well be that we find something completely unexpected in the course 
of other studies that will finally lead to the long-waited understanding 
of this exceptional hydrodynamical phenomenon. 

\vspace*{-2mm}
\section*{Acknowledgements}
\vspace*{-2mm}
My life-long appreciation and gratitude go out to my mentor and long-time 
friend Wojtek who introduced me into the exciting field of nonlinear stellar 
pulsations and gave me his support and encouragement over the years.

\end{document}